\documentstyle[amssymb,amsmath,12pt]{article}

\def\slash#1{\setbox0=
\hbox{$#1$}#1\hskip-\wd0\hbox to\wd0{\hss\sl/\/\hss}}

\begin{document}


%
%

\begin{center}
\bf Abrikosov Vortex and Branes\\[1cm]


\rm S. Randjbar-Daemi\\
International Centre for Theoretical Physics,\\
 34100 Trieste, Italy
\end{center}
\bigskip

\begin{abstract}

 In this contribution in honor of A.P. Balchandran  we  give a brief
description of application of topological solutions in field
theory, a subject which has been central to many of
 Balachandran's important contributions to physics.

\end{abstract}

\section{Introduction}

Beautiful mathematics  and its application in physics is one of
the main characteristics of Balchandran's work.  The application
of topological ideas in physics permeates throughout most of Bal's
papers. It is a pleasure to read them and I personally have always
found them enlightening. A collection of some of these  can be
found in the monograph.~\cite{bal}

 Topological configurations have played important role in our
understanding of condensed matter as well as particle physics. In
high energy physics they appear as finite energy or finite (
Euclidean-) action solutions of non linear classical field
equations while  in string theory they are  p-brane configurations
with non zero charges defined by space integrals of differential
forms of appropriate ranks.~\cite{Polyak},~\cite{Pol} Magnetic
systems are understood in terms of domain formation, phase
transitions in two dimensional systems with continuous  symmetries
are explained in terms of formation of vortices.~\cite{KT} In
particle physics our only understanding of the famous $U(1)$
problem is in terms of the instanton solution.~\cite{'thooft2}
Formation of vortices, domain walls and magnetic monopoles have
relevance in cosmology.~\cite{Vilen} Also one of the ideas which
recently acquired considerable interest is the possibility that
our 4-dimensional space time may be a topological defect of a
3-brane type extending in a higher dimensional space
time.~\cite{Rub.Sh} Unlike the classical Kaluza Klein theory where
one assumed that the extra dimensions should be small and cover a
compact manifold in this paradigm the extra dimensions can be
large and non compact.~\cite{Rub.Sh2},~\cite{RW} This freedom
allows a reformulation of one of the basic questions of
fundamental physics, namely, why the planck mass is so much larger
than the electroweak symmetry breaking scale.~\cite{RandSund}.
 In this contribution we shall give a very brief review of some of these
ideas. In section 2 we start from the Kosterliz-Thouless
description of phase transitions in two dimensional systems with
continuous symmetries in terms of vortex formations and go on to
the vortex solutions in 3 and higher dimensions. In section 3 we
discuss very briefly the formation of walls, and fermion
localization on them. In section 4 we mention some examples in
string theory.
\section{Vortices in 1+1 and higher dimensions}
\subsection{1+1 dimensions}

Consider a field $\theta(x)$ in $1+1$ dimensions which takes its
values in a circle. The Hamiltonian is

\begin{equation}
H= J/2\int d^2x \nabla_i\theta\nabla_i\theta
\end{equation}

The non triviality of this system is related to the fact that
$\theta(x)$ ia an angle. Shifting $\theta (x)$ by an $x-$
independent angle does not change $H$. Thus there is a continuous
$U(1)$ symmetry in this problem. It is not hard to show that this
symmetry can not be broken spontaneously in $1+1$ dimensions. The
low temperature fluctuations are spin waves with power law
correlations. Therefore there can not be a  phase transitions
which is triggered by spontaneous symmetry breaking. There is a
continuous line of critical points with temperature dependent
exponent.  It was shown by Kosterlitz and Thouless that actually a
sharply defined transition temperature does exist and  the phase
transition proceeds by the formation of vortices.~\cite{KT} These
configurations are defined by those $\theta(x)$ for which $\nabla
\theta(x)$ are multiple valued, viz.$\int_{C} dl_i \nabla_i\theta
= 2n\pi n$,
 where $C$ is a contour which encircles the origin and $n$ is any integer.
Away from the origin one can write ( we take n=1
case)$\nabla_i\theta(x)= (0, 1/r)$.The energy of this
configuration will be $E=\pi J ln (L/a)$. Here $a$ is a short
distance cut off, the lattice spacing in spin systems to which
these considerations are normally applied, and $L$ is the size of
the system. Since the center of the vortex can be located at any
one of the $(L/a)^2$ points on the lattice the entropy for the
creation of a vortex will be $ S= ln(L/a)^2$. The free energy
associated with the creation of a vortex can easily be obtained to
be $F= E- TS = (\pi J- 2T) ln (L/a)$. There is thus a critical
temperature $ T_c= \pi J/2$ above which the quasi order of the
spin waves are unstable against formation of single vortices.
Below $T_c$ we have only vortex-antivortex pair. As $T$ increases
the size of vortex-antivortex pair increases and above $T_c$ the
pair becomes unstable. This destabilize the quasi order of the low
$T$ spin waves and makes the correlations to decay
 exponentially. This is the KT description of
phase transitions in two dimensional spin systems with continuous
symmetries. Note that the formation of vortices in this system is
entirely due to topology, namely, the fact that $\theta$ takes its
values on a $S^1$ and that $\pi_1 (S^1) =Z$.

 \subsection{Higher than 2-dimensions}

For the application in particle physics and cosmology the
interesting case arises when we couple gauge fields to scalars in
some representation of the gauge group. The scalars are assumed to
have a potential which allows spontaneous breaking of the gauge
symmetry $G$ to one of its subgroups $H$. It is the topology of
the vacuum manifold $G/H$ which plays the key role in ensuring the
existence of topologically non trivial configurations. The
simplest case is to have a U(1)gauge theory coupled to a complex
scalar field $\Phi$. In 3 space dimensions, provided that the
charge of $\Phi$ is twice that of the electron, this will give the
Ginzburg-Landau description of the BCS theory of
superconductivity. The spontaneous breaking of the $U(1)$ symmetry
gives rise to a finite penetration depth. This so called Meissner
effect can also be seen in $2+1$ dimensional gauge theories of
Chern Simons type without a need for spontaneous symmetry
breaking. However, unlike the Higgs induced order which disappears
at some finite temperature, the CS induced Meissner effect does
not seem to vanish at any $T_c$. The penetration depth remains
finite at all temperature, at least in perturbation theory.
~\cite{ChernSuper} Abrikozov demonstrated
 the
formation of vortices in the LG model and discussed their physical
relevance for type II superconductivity.~\cite{ab} The same system
was then studied by Nielsen and Olesen in the context of $3+1$
dimensional scalar electrodynamics.~\cite{NO} Here the vortices
will appear as static solutions. Their formation was  thought to
be relevant for quark confinement. Another very interesting
consequence of this type of solution is the localization of
fermionic zero modes to the core of the vortex. The fact that this
does indeed happen was  shown in Ref.~\cite{JR}. We shall come
back to this issue in the following section and discuss it in some
detail in 6 dimensions. If the localized fermions are electrically
charged they can lead to the formation of superconduction cosmic
strings.~\cite{wit}

One can proceed to higher than $4$ dimensions and apply these
ideas to
 brane world scenarios.  The simplest model for our world as a
brane extending in a higher dimensional space time is perhaps the
one constructed by Rubakov and Shaposhnikov.~\cite{Rub.Sh}  In the
absence of gravity this model consists of a scalar field with an
appropriate self coupling to allow for the spontaneous breaking of
a $Z_2$ symmetry. Denote the fifth coordinate by $r$ and the first
four coordinates by $x^\mu$. The r-dependent (and $x^\mu$
independent) kink solution for the scalar field will be seen as a
3-brane localized along the $r$ axis. Its thickness will be given
by the thickness of the kink, which in turn is determined by the
Compton wavelength of the scalar particle. The interesting result
of Rubakov and Shaposhnikov was to show that if we study the
solutions of the 5-dimensional Dirac equation in the kink
background we shall find two, 4-dimensional chiral configurations.
Only one of them has a normalizable  kinetic energy and is
localized to the wall. This result should be contrasted with the
Kaluza Klein ideas according to which the extra dimensions are
assumed to be compact contrary to the non compact r coordinate.
Also to obtain chiral fermions one needed to start from an even
dimensional theory with fundamental gauge fields coupled to chiral
fermions in the higher dimensional space time.~\cite{RSSMon}
~\cite{Witten}   If we couple a charged scalar to the bosonic part
of the same 6-dimensional system we can search for vortex
solution. Such a solution has indeed been found and has been
interpreted as a 4-dimensional universe sitting at the core of the
vortex.~\cite{MGivann1} In order for the 4-dimensional physics to
reside at the core  one needs to find physical mechanisms which
localize the higher-dimensional fields to the 3-dimensional brane
residing at the core  and whose world volume is our 4-dimensional
space-time. In general the localization of fermions is relatively
straightforward in the presence of warped geometry and  non
trivial Yang-Mills backgrounds.~\cite{RsMish1}In the specific
vortex background the relevant equations can be reduced to a form
which agrees with the familiar case of $3+1$ dimensional problem
near the core of the vortex.~\cite{JR} Far from the core they are
different.~\cite{QED} It turns out that the  gauge field
fluctuations also do localize. However, there is no mass gap in
the spectrum of the localized gauge fields to separate them from
the bulk excitations. The fermion spectrum on the other hand does
have a mass gap. In the absence of such a gap for the gauge fields
it is unclear how one can neglect the bulk modes and consider only
the localized modes. Many aspects of the localization problem
recently has been a subject of an intense study.~\cite{Mass}
 Allowing for non compact internal
dimensions raised some early hopes for a solution of the
cosmological constant problem.~\cite{Rub.Sh2}'~\cite{RW}  As
mentioned before, recently the idea of a non compact extra
dimension raised the hope of reformulating the well known
hierarchy problem.~\cite{RandSund}Another reformulation of the
same problem is in the context of a large but compact extra
dimension ~\cite{ADDD}, although the idea of a large compact extra
dimensions is not new.
\subsection{Vortices in 6-dimensions}

Now we go a little more into the detail of the 6-dimensional
vortex solution. We are interested in a warped geometry for the
space-time and vortex configuration for the gauge-scalar system.
The background geometry is defined by the metric,

$$
ds^2 = e^A(r)\eta_{\mu\nu}dx^\mu dx^\nu +dr^2 + e^B(r) a^2
d\theta^2
$$
where $\mu, \nu =0,1,2,3$ and $a$ is the radius of $S^1$ covered
by the $\theta$ coordinate. The ansatz for the gauge field $A_M$
and the scalar $\Phi$  will be a Nielsen-Olesen vortex solution. $
\Phi= f(r)e^{in\theta} $
 and
$ aeA= (P(r)-n) d\theta. $
 The functions $f(r)$ and
$P(r)$ satisfy the following boundary conditions
\renewcommand{\theequation}{2.3}
\begin{equation}
\begin{array}{ll}
f(0)=0,\qquad\quad &   f(\infty)=f_0\\[2mm]
P(0)=n,& P(\infty)=0
\end{array}
\end{equation}

In Ref.~\cite{MGivann1} it was shown that the above ansatz solves
the bosonic background equations. There are solutions with
different boundary conditions for the metrical fields $A$ and $B$.
The boundary conditions of interest for us which localize fields
of spin $ 0, 1/2 \;\mbox{and}\; 1$ are, as $r\rightarrow 0$, given
by $ A(r)\rightarrow 0 \; \;\mbox{and} \;\; B(r) \rightarrow
2\;\mbox{ln}(r/a) $, while as $r\rightarrow \infty$ we should
demand $A(r)= B(r)\rightarrow -2cr$, where $c$ is a positive
constant.

As has been explained in  Ref.~\cite{QED} in order to find the
correct anzatz for the $U(1)$ gauge field we need to identify the
unbroken symmetry group which leaves the background geometry as
well as the vortex configuration invariant. This leads us to the
identification

\renewcommand{\theequation}{2.6}
\begin{equation}
V_i ^{(1)} = \frac{1}{ae} P(r) W_i (x), \;\;\;
 and \;\;\; h_{i\theta}
=e^B W_i (x)
\end{equation}
where $W_i$ is a function of $x^\mu$ only, and i=1,2 are the
transverse directions in $D=4$. It is the  gauge field of the
unbroken $U(1)$. This anstaz leads to the following kinetic energy
for $W_i$ fields~\cite{QED}
\renewcommand{\theequation}{2.7}
\begin{equation}
S(W_\mu)= {1\over 2a^2e^2} \int d^6x\ e^{{1\over 2}B}
\left(P^2(r)+{a^2e^2\over \kappa^2}e^B\right)
\partial_\nu W_i \partial^\nu W_i
\end{equation}

The r-integral is converging quite rapidly and we obtain a
normalizable effective action for $W_i$.

Note that in the above formula we have the bilinear part of the
Maxwell action in the light cone gauge . For that reason only the
transverse components of the gauge field appear in the action. In
the non covariant gauge used in this calculation the light cone
components $V_{-} $ of the gauge field fluctuations and $h_{-M}$
of the gravitational fluctuations are set to zero. The equations
of motions for the remaining longitudinal components $V_{+}$ and
$h_{+M}$ then are determined in terms of the transverse components
$V_i$ and $h_{ij}$, etc. In this way the redundant components
disappear completely from the formalism. This simplifies the
spectrum analysis of the massless modes .~\cite{QED}

 \section{Fermions}
Fermion localization on defects has many interesting physical
implications.  Here we consider three examples. The first is
basically a two dimensional phenomena which results from the fact
that the $1+3$ dimensional Dirac equation in the background of a
Nielsen-Olesesn string has localized zero modes.~\cite{JR} These
modes are massless and propagate with the speed of light along the
string. It was shown by Witten that, if charged, they can lead to
superconducting cosmic strings.~\cite{wit} Their presence in grand
unifying theories have significant cosmological implications.

The second example concerns the existence of fermion zero modes in
D=6 in a vortex background and warped geometry. Their presence
leads to D=4 QED near the core of the vortex. ~\cite{QED}. In the
absence of gravity the only way to find normalizable solutions of
Dirac equation in the background of a vortex in $1+3$ dimensions
is to introduce a Yukawa coupling as it was originally done in
~Re.\cite{JR}. In the presence of gravity and warped geometry,
however, there exist normalizable zero mode solutions even in the
absence of Yukawa couplings. For the consistency of the starting
theory one needs to make the six dimensional theory free from
chiral anomalies. This can  be done. In fact there exist anomaly
free supergravity models in $D=6$ with ungauged ~\cite{GSW} or
gauged~\cite{RSSS} R-symmetries. Furthermore it has been shown
that the spectrum of localized fermions is separated by a mass gap
from the bulk fermions. Thus unlike the vector bosons they have a
very clean physical interpretation in the context of brane world
scenarios.

To study the problem of fermion mass gap we need only to consider
the large $r$ limit, for which, the Yukawa  and Majorana mass
terms in the fermionic part of the action  can be neglected, and
the equations for bulk modes read (we consider only $4_+$
six-dimensional fermion, the case of $4_-$ is treated in full
analogy) :

\begin{eqnarray}
{\rm e}^{+{\rm i}\theta} \left(\partial_r+{{\rm i}\over a} {\rm
e}^{-B/2}\partial_\theta\right)\psi^L &=&~ m_f e^{-\frac{1}{2} A
-2\int^r e^{-\frac{B}{2}}e_L A_\theta d\rho}
\psi^R~,\\
\nonumber {\rm e}^{-{\rm i}\theta} \left(\partial_r-{{\rm i}\over
a} {\rm e}^{-B/2}\partial_\theta\right)\psi^R &=& - m_f
e^{-\frac{1}{2} A +2\int^r e^{-\frac{B}{2}}e_L A_\theta d\rho}
\psi^L~.
\end{eqnarray}
The ansatz $\psi_L=e^{im\theta}\psi_L^m$,
$\psi_R=e^{i(m+1)\theta}\psi_L^m$ removes the angular dependence
and leads, for large $r$, to the following  asymptotic
 solutions
\begin{equation}
\psi_L \to \frac{1}{\sqrt{z}} \exp
\left(-\frac{1-2ne_L/e}{2am_f}z\right) \left(C_1 \sin(S z) + C_2
\cos(Sz)\right)~,
\end{equation}
where $z=\frac{m_f}{c} e^{cr}$ and $ S^2=1-
\frac{(m-ne_L/e+\frac{1}{2})^2}{(am_f)^2} ~ \label{root} $. From
this we immediately see that the charged bulk fermions are
massive, $m_f >\frac{1}{a}|m-\frac{ne_L}{e} +\frac{1}{2}|$, since
the combination of $m$ and $n$ in $S^2$ is exactly the electric
charge. There are of course also charged localized fermions which
are massless.

For our third example we mention only in passing how fermion
localization on a domain wall has been employed to give a non
perturbative definition of chiral gauge theories, i.e a way of
defining chiral fermions on the lattice without running into the
doubling problem.

 Let us consider a 5-dimensional space-time. Assume that there is a
4-dimensional wall sitting at some point on the time axis, as
opposed to the spatial $r$-axis discussed before. This system has
a time dependent mass term in the  Hamiltonian. The time
dependence is simply given by $sign(t)$, which is $+1$ for
positive $t$ and $-1$ for negative $t$. There are a finite number
of chiral fermion modes localized to the wall. Kaplan made the
interesting suggestion that this fact may be used to solve the
long standing problem of defining Euclidean lattice chiral gauge
theories by identifying the wall with the 4-dimensional Euclidean
space.~\cite{Kaplan} Narayanan and Neuberger removed the
shortcomings of this idea and turned  it into a formalism for a
non perturbative definition of chiral gauge theories. They gave a
formula for the chiral Dirac operator as the overlap between two
vectors in the Hilbert space of this set up.~\cite{NN} These
vectors are denoted by $\vert A-\rangle$ and $\vert A+\rangle$ in
the notation of Re \cite{Randjbar-Daemi:1997iz}, where $A$ refers
to the gauge field. The chiral Dirac determinant is then
identified with their inner product and chiral anomalies have
their root in the phase of this complex functional. Several
aspects of this prescription has been examined in detail.  In
particular it has been shown that the formalism correctly
reproduces all one loop amplitudes for slowly varying external
gauge fields, including chiral gauge and gravitational anomalies
\cite{Randjbar-Daemi:1997iz}.

\section{String Theory} Superstring theory allows for
phenomena which may not exist or not easy to observe in field
theory. For example it is possible that the topology of a
Calabi-Yau manifold undergoes a change without creating a
singularity in space time physics. This happens by passing through
a conifold point in the moduli space of CY
manifolds.~\cite{Candelas}Starting with a smooth CY with Euler
number $\chi$ one can shrink $N$ $S^3$'s to single points, thereby
obtaining a singular manifold with the Euler number of $\chi +N$
sitting at a conifold singularity in the moduli space. In the next
step one replaces each singular point with a $S^2$ to obtain a
smooth CY with an Euler number of $\chi+2N$.

 A beautiful interpretation of this conifold transition is as
follows: Consider a $D_3$ brane wrapped around a $S^3$ in
compactification  space of  the type IIB string on some $CY_3$
three folds. The low energy dynamics of the  moduli fields is
described by a $\sigma$ model targeted on $CY_3$. These fields in
their time evolution encounter a conifold singularity which
corresponds to vanishing $S^3$'s. The vanishing volume of the
wrapped $D_3$ brane in its turn gives rise to the emergence of the
massless states in the low energy spectrum. Thus physically, the
appearance of singularity is equivalent to the emergence of new
massless multiplets in the 4-dimensional spectrum.~\cite{strom}

Conifolds enter many other discussions in string theory. One very
interesting case is the demonstration by Witten that the free
energy of a $U(N)$ Chern-Simons theory on $S^3$ equals the free
energy of topological open strings targeted on the cotangent
bundle of $S^3$, which is a deformed conifold.~\cite{witten}The
free energy of the CS theory can be formally expanded as $
E=\sum_{g=0,h=1} C_{g,h} N^{2g-2} \lambda^{2g-2+h}$, where
$\lambda =\frac{k}{N+k}$. According to Witten each term in the
above sum can be interpreted as the free energy of an open string
diagram on a world sheet of genus $g$ and $h$ boundaries.Assume we
perform the sum over $h$. The result should be an expression of
the form $ E=\sum_{g=0}^\infty  N^{2g-2} F(\lambda)$ which looks
like the topological sum over all closed string world sheet, each
term representing the contribution of a genus $g$ surface. The
string coupling constant $g_s$ is clearly proportional to $1/N$
and vanishes as $N$ goes to infinity. In fact one can calculate
$F_g$ for all values of $g$ and show that each $F_g$ is identical
to the free energy of a closed string targeted on a small
resolution of a $CY_3$. Thus it seems that summing over the
boundaries of the world sheet  generates a transition from a $S^3$
deformed conifold through a singular conifold to a small
resolution.~\cite{GV}

\section*{Acknowledgments}

I would like to thank Misha Shaposhnikov for collaboration and
many useful discussions. I would also like to thank the organizers
of the Balfest for inviting me to make this contribution. The work
reported here has been partially supported by the EU fund,
CT-200-00148.

\end{document}